\documentclass[usenatbib,fleqn,useAMS,twocolumn]{mnras}
\usepackage[dvipdfmx]{graphicx}
\usepackage[dvipdfmx]{color}
\usepackage{amsmath,amssymb}
\usepackage{epstopdf}
\usepackage{grffile}
\usepackage{times}
\usepackage{url}
\usepackage{bm}
\usepackage{xcolor}

\newcommand{\simgt}{\lower.5ex\hbox{$\; \buildrel > \over \sim \;$}}
\newcommand{\simlt}{\lower.5ex\hbox{$\; \buildrel < \over \sim \;$}}

\newcommand{\himpc}{{\hbox {$~h^{-1}$}{\rm ~Mpc}}}
\newcommand{\higpc}{{\hbox {$~h^{-1}$}{\rm ~Gpc}}}

\newcommand{\himsun}{{\hbox {$~h^{-1}$}{M_\odot}}}

\newcommand{\vx}{\mathbf{x}}

\newcommand{\vk}{\mathbf{k}}

\newcommand{\vv}{\mathbf{v}}
\newcommand{\vecr}{\mathbf{r}}

\newcommand{\be}{\begin{equation}}
\newcommand{\ee}{\end{equation}}
\newcommand{\bey}{\begin{eqnarray}}
\newcommand{\eey}{\end{eqnarray}}
\newcommand{\bea}{\begin{align}}
\newcommand{\eea}{\end{align}}

\newcommand{\wt}{\widetilde}

\newcommand{\nn}{\nonumber}

\newcommand{\regpt}{R{\footnotesize EG}PT }
\newcommand{\camb}{\texttt{\footnotesize CAMB} }
\newcommand{\rockstar}{\texttt{\footnotesize ROCKSTAR} }
\newcommand{\darkquest}{\texttt{D{\footnotesize ARK} Q{\footnotesize UEST} }}

\begin{document}
\title[Anisotropic correlations of halo ellipticities]
{
Testing tidal alignment models for anisotropic correlations of halo ellipticities with $N$-body simulations
}

\author[T. Okumura, A. Taruya and T. Nishimichi]{
\parbox{\textwidth}{
Teppei Okumura$^{1,2}$\thanks{tokumura@asiaa.sinica.edu.tw}, Atsushi Taruya$^{3,2}$ and Takahiro Nishimichi$^{3,2}$}
\vspace*{4pt} \\
$^{1}$ Academia Sinica Institute of Astronomy and Astrophysics (ASIAA),
No. 1, Section 4, Roosevelt Road, Taipei 10617, Taiwan \\
$^{2}$ Kavli Institute for the Physics and Mathematics of the Universe (WPI), 
UTIAS, The University of Tokyo, Kashiwa, Chiba 277-8583, Japan \\
$^{3}$ Center for Gravitational Physics, Yukawa Institute for Theoretical Physics, Kyoto University, Kyoto 606-8502, Japan
}

\date{\today} 
\pagerange{\pageref{firstpage}--\pageref{lastpage}} \pubyear{2020}

\maketitle
\label{firstpage}

\begin{abstract}
There is a growing interest of using the intrinsic alignment (IA) of galaxy images as a tool to extract cosmological information complimentary to galaxy clustering analysis. Recently, Okumura \& Taruya derived useful formulas for the intrinsic ellipticity--ellipticity correlation, the gravitational shear--intrinsic ellipticity correlation, and the velocity--intrinsic ellipticity correlation functions based on the linear alignment (LA) model. In this paper, using large-volume $N$-body simulations, we measure these alignment statistics in real and redshift space and compare them to the LA and nonlinear alignment  model predictions. We find that anisotropic features of baryon acoustic oscillations in the IA statistics can be accurately predicted by our models. The anisotropy due to redshift-space distortions (RSDs) is also well described in the large-scale limit. Our results indicate that one can extract the cosmological information encoded in the IA through the Alcock-Paczynski and RSD effects.
 \end{abstract}


\section{Introduction} \label{sec:intro}

The intrinsic alignment (IA) of galaxy images \citep{Heavens:2000,Croft:2000,Lee:2000,Catelan:2001,Crittenden:2001,Hirata:2004} \cite[See][for reviews]{Troxel:2015,Joachimi:2015,Kirk:2015,Kiessling:2015} is known to contaminate cosmological parameter estimations from weak lensing surveys, by $5-10\% $ \citep{Mandelbaum:2006,Hirata:2007,Okumura:2009,Joachimi:2011,Singh:2015}. Precise modeling of the intrinsic alignments (IAs) is thus crucial for the cosmic shear to be a useful cosmological tool \citep{Jing:2002,Smith:2005,Bridle:2007,Okumura:2009a,Zhang:2010,Schneider:2010,Blazek:2011,Blazek:2015,Xia:2017,Codis:2018,Yao:2019,Yao:2019a,Vlah:2019,Blazek:2019,Okumura:2019}. 
On the other hand, less focus has been drawn to the cosmological information IAs encode. \cite{Schmidt:2012} claimed that gravitational waves from inflation could be detected in intrinsic ellipticities of galaxies. \cite{Faltenbacher:2012} and \cite{Chisari:2013} argued the detectability of baryon acoustic oscillations \citep[BAOs, ][]{Sunyaev:1970,Peebles:1970,Hu:1996,Eisenstein:2005} in statistics of the IAs. Recently there are more and more follow-up studies which utilize the IA as cosmological probes \citep{Chisari:2014,Schmidt:2015,Chisari:2016,Kogai:2018,Yu:2019}. 

In these studies, however, only the 1-dimensional statistics, e.g., the angular, projected and monopole correlation functions, have been used to quantify the IAs. By construction, anisotropies arise in the alignment statistics even in real space because observed galaxy images are the projection along the observer's line of sight. Thus, analyzing angular correlation statistics is not sufficient to extract the full cosmological information encoded in the IAs. \cite{Okumura:2019a} presented analytic formulation of alignment statistics in real and redshift space based on the linear alignment (LA) model \citep[][see Sec. \ref{sec:theory} below]{Catelan:2001,Hirata:2004} and derived the full angular dependences for the statistics. The derived formulas enable us to analyze anisotropies of BAOs in the IAs and extract the maximum cosmological information using the Alcock-Paczynski effect \citep{Alcock:1979} and redshift-space distortions \cite[RSDs, ][]{Kaiser:1987,Hamilton:1992} from ongoing and upcoming galaxy redshift surveys, such as the Sloan Digital Sky Survey III (SDSS-III) Baryon Oscillation Spectroscopic Survey \citep[BOSS,][]{Eisenstein:2011}, the Dark Energy Spectroscopic Instrument \citep[DESI,][]{DESI-Collaboration:2016}, and the Subaru Prime Focus Spectrograph \citep[PFS,][]{Takada:2014}. See \cite{Hirata:2009} and \cite{Martens:2018} for the studies of how RSD measurements are affected by the IA. 

In this paper, we test the tidal alignment model predictions for IA statistics derived in \cite{Okumura:2019a} in real and redshift space on large scales, using a large set of cosmological $N$-body simulations. In a companion paper, \cite{Taruya:2019}, we present a Fisher matrix analysis to show that combining the IA statistics to the conventional clustering statistics significantly tightens cosmological constraints. 
The rest of this paper is organized as follows.
In section \ref{sec:theory}, we briefly review the statistics of IAs used in this paper and present the formulas of the IA statistics based on the LA model. 
Section \ref{sec:simulation} describes $N$-body simulations and estimators to measure the IA statistics in the simulations.
In section \ref{sec:result} we compare the LA model and its extension, nonlinear alignment (NLA) model, for the IA statistics to the $N$-body measurements.
Our conclusions are given in section \ref{sec:conclusion}. 

\section{Intrinsic alignment statistics} \label{sec:theory}
In this section we briefly describe the alignment statistics and the linear tidal alignment model used in this paper. See \cite{Okumura:2019} and \cite{Okumura:2019a} for the more detailed description. 

\subsection{Definitions}\label{sec:definition}

The intrinsic ellipticity of galaxies or halos, defined on the celestial sphere as an observable, is characterized by the two-component quantity,
\begin{align}
\gamma_{(+,\times)}(\vx)=\frac{1-(\beta/\alpha)^2}{1+(\beta/\alpha)^2}(\cos(2\theta),\sin(2\theta)), \label{eq:gamma_pc}
\end{align}
where $\theta$, defined on the plane normal to the line-of-sight direction, represents the angle between the major axis projected onto the celestial sphere and the projected separation vector pointing to another object, and $\beta/\alpha$ is the minor-to-major-axis ratio for the projected shape. Throughout this paper, we set $\beta/\alpha=0$ for simplicity, which corresponds to the assumption that a galaxy shape is a line along its major axis \citep[e.g.,][]{Okumura:2009,Okumura:2009a}.

The intrinsic-intrinsic ellipticity correlations (II correlations) have four components, and one of the four, $\xi_{++}$, is defined as \citep{Heavens:2000, Croft:2000}
\be
1+\xi_{++}(\vecr)=\left\langle[1+\delta_h(\vx_1)][1+\delta_h(\vx_2)]\gamma_+(\vx_1)\gamma_+(\vx_2)\right\rangle,  \label{eq:ii}
\ee
where $\vecr=\vx_2-\vx_1$ and $\delta_h(\vx)$ is the galaxy/halo overdensity. We use the subscript $h$ because in the following sections halos are used as a tracer of the ellipticity field in $N$-body simulations. The other components, such as $\xi_{\times\times}$ and $\xi_{+\times}$, are defined in the same way by replacing two and one $\gamma_+$ in Equation (\ref{eq:ii}) with $\gamma_\times$, respectively. By combining $\xi_{++}$ and $\xi_{\times\times}$, we can also define $\xi_\pm(\vecr)$ as
\be 
\xi_\pm(\vecr) = \xi_{++}(\vecr) \pm \xi_{\times\times}(\vecr). \label{eq:iipm}
\ee
In this paper, the II correlation functions refer to these two functions, $\xi_\pm$. 
The cross correlation functions of density and ellipticity fields, namely gravitational shear--intrinsic ellipticity correlations (GI correlations), are defined as \citep{Hirata:2004}
\be
1+\xi_{hi}(\vecr)=\left\langle[1+\delta_h(\vx_1)][1+\delta_h(\vx_2)]\gamma_i(\vx_2)\right\rangle. \label{eq:gi}
\ee
where $i=\{+,\times \}$.
We also consider the velocity alignment statistic, the density-weighted velocity-intrinsic ellipticity (VI) correlation \citep{Okumura:2019} as,
\begin{align}
\xi_{vi}(\vecr)=\left\langle[1+\delta_h(\vx_1)][1+\delta_h(\vx_2)] v_\parallel(\vx_1)\gamma_i (\vx_2) \right \rangle, \label{eq:vi}
\end{align}
where $i=\{+,\times\}$ and $v_\parallel$ denotes the line-of-sight component of the velocity field, $v_\parallel(\vx)\equiv \vv(\vx)\cdot \hat{\vx}$ (hat means a unit vector). 

All the statistics above are anisotropic even in real space because observable shapes of galaxies are the line-of-sight projection. Moreover, RSDs induce further anisotropies particularly to the GI correlation function. Thus, we will present these statistics as a function of separations perpendicular ($r_\perp$) and parallel ($r_\parallel$) to the line-of-sight direction, $X(\vecr)=X(r_\perp,r_\parallel)$, where $X$ is any of the statistics introduced above. When we want to make a detailed comparison between the measurement from $N$-body simulations and the corresponding model prediction, we will further consider the multipole expansions of the correlation functions \citep{Hamilton:1992}:
\begin{align}
X_{{\ell}}(r)=\frac{2\ell+1}{2}\int^1_{-1} d\mu X(\vecr){\cal P}_\ell(\mu), 
\end{align}
where $r = |\vecr| = (r_\perp^2 + r_\parallel^2)^{1/2}$, ${\cal P}_\ell$ is the Legendre polynomials, and $\mu$ is the directional cosine between the vector $\vecr$ and the line-of-sight direction $\vx$, $\mu = r_\parallel / r$ (Note that $\mu \neq \cos{\theta}$). Throughout this paper, we assume the distant-observer approximation so that $\hat{\vx_1}=\hat{\vx_2}\equiv \hat{\vx}$. 

\subsection{Linear alignment model in three dimension}

In this paper we use the linear alignment (LA) model \citep{Catelan:2001,Hirata:2004} and its extension, the nonlinear alignment (NLA) model \citep{Bridle:2007} to predict the IA statistics measured from $N$-body simulations. The LA model has been well studied observationally and numerically and is known to predict the monopole component of the IA statistics in the large-scale limit \citep{Blazek:2011,Schneider:2012,Blazek:2015,Okumura:2019}. In this paper we will test the LA model for the IA statistics with the anisotropic features derived in \cite{Okumura:2019a}. 

The LA model assumes the linear relation between the ellipticity field and the tidal gravitational field,
\be
\gamma_{(+,\times)}(\vx)= -\frac{C_1}{4\pi G}\left( \nabla_x^2-\nabla_y^2, 2 \nabla_x\nabla_y \right) \Psi_P(\vx), 
\label{eq:gamma_la}
\ee
where $\Psi_P$ is the gravitational potential related to the mass density field $\delta$ through $\Psi_P(\vk)= -4\pi G \left[ \bar{\rho}(a)a^2 / \bar{D}(a) \right]k^{-2}\delta(\vk) $ in Fourier space, $G$ is the gravitational constant, $\bar{\rho}$ is the mean mass density of the Universe, $\bar{D}(a)\propto D(a)/a $ where $D(a)$ is the linear growth factor with $a$ being the scale factor, and the parameter $C_1$ characterizes the strength of the IAs. The observed ellipticity field is density-weighted, $[1+\delta_h(\vx)]\gamma_{(+,\times)}(\vx)$. However, the density-weighting term $\delta_h(\vx)\gamma(\vx)$ is sub-dominant on large scales and is usually ignored. We do not consider this term because we are interested in the large-scale behaviors. In Fourier space, Equation (\ref{eq:gamma_la}) becomes
\begin{align}
\gamma_{(+,\times)} (\vk) =  -\wt{C}_1(a) 
\frac{\left(  {k}_{x}^{2}- {k}_{y}^{2}, 2{k}_x{k}_y \right)}{k^2}\delta(\vk),
\end{align}
where $\wt{C}_1(a)\equiv a^2 C_1\bar{\rho}(a)/\bar{D}(a)$.

In the following paragraphs, we use the expression introduced in \cite{Okumura:2019a}:
\be
\Xi_{XY,\ell}^{(n)}(r) = (aHf)^n\int^{\infty}_0 \frac{k^{2-n}dk}{2\pi^2}P_{XY}(k) j_\ell(kr), \label{eq:xi_formula}
\ee
where $XY=\{ \delta\delta, \delta\Theta\}$, $\Theta$ is the velocity-divergence field defined by $\Theta(\vx)=-\nabla \cdot \vv/(aHf)$, $H(a)$ is the Hubble parameter and $f(a)$ is the linear growth rate, given by $f(a)\equiv d\ln D/d\ln a$. The quantities $P_{\delta\delta}$ and $P_{\delta\Theta}$ are the auto power spectrum of density and the cross power spectrum of density and velocity divergence, respectively. In the linear theory limit, we have $P_{\delta\delta}=P_{\delta\Theta}$. 

The GI correlation function in real space is given by
\be
\xi_{h+ }^R(\vecr) 
=\wt{C}_1 b_h 
(1-\mu^2) 
\Xi_{\delta\delta,2}^{(0)}(r),
\label{eq:gi_la_1d}
\ee
where the superscript $R$ denotes a quantity defined in real space, while the superscript $S$ which appears below denotes a quantity defined in redshift space, and $b_h$ is the linear bias parameter of halos. Strictly, $\xi_{h+}^R$ contains a prefactor of $\cos{(2\phi)}$, where $\phi$ is the azimuthal angle of the projected separation vector on the celestial sphere. However, here and in the following we set $\phi=0$ and omits the $\phi$-dependence \cite[see][for the complete forms]{Okumura:2019a}. One can easily show that the multipole components which become non-zero are only monopole and quadrupole, given by 
\be
\xi_{h+,0 }^R(r) =-\xi_{h+,2 }^R(r)  =\frac{2}{3}\wt{C}_1 b_h\Xi_{\delta\delta,2}^{(0)}(r). \label{eq:gi_la_l}
\ee
In order to derive the expression in redshift space, we need to specify the model of RSDs. We consider the simple Kaiser RSD model \citep{Kaiser:1987}, in which the redshift-space density field, $\xi_h^S$, is expressed in Fourier space as
\be
\delta_h^S(\vk)=\delta_h^R(k)+f\,(k_z/k)^2\Theta(k),
\ee
where $\Theta(k)$ is the Fourier transform of the velocity divergence. The linear growth rate $f$ in the RSD formula is used to constrain modified gravity models \citep{Wang:1998,Linder:2005,Guzzo:2008,Okumura:2016}. The GI correlation function in redshift space is then obtained as 
\begin{align}
\xi_{h+ }^S(\vecr) 
&=\xi_{h+ }^R(\vecr) 
+\frac{1}{7}\wt{C}_1 f 
\left(1-\mu^2\right) \nn \\
&\times\left[\Xi_{\delta\Theta,2}^{(0)}(r)-\left(7\mu^2-1\right)  \Xi_{\delta\Theta,4}^{(0)}(r)\right].
\label{eq:gi_la_s_1d}
\end{align}
Compared to the real-space correlations, not only the monopole and quadrupole but also hexadecapole moments are non-zero in the LA model:
\begin{align}
\xi_{h+,0 }^S(r) 
&=\xi_{h+,0 }^R(r) 
+\frac{2}{105}\, \wt{C}_1 f 
\left[5\,\Xi_{\delta\Theta,2}^{(0)}(r)- 2\,\Xi_{\delta\Theta,4}^{(0)}(r) \right], \\
\xi_{h+,2 }^S(r) 
&=\xi_{h+,2 }^R(r) 
-\frac{2}{21}\, \wt{C}_1\, f \, 
\left[\,\Xi_{\delta\Theta,2}^{(0)}(r)+ 2\,\Xi_{\delta\Theta,4}^{(0)}(r)\right], \\
\xi_{h+,4 }^S(r) 
&=
\frac{8}{35}\, \wt{C}_1\, f \,
\Xi_{\delta\Theta,4}^{(0)}(r).
\end{align}

The II correlation functions, $\xi_\pm^R(\vecr)$, are expressed as 
\begin{align}
\xi_{+}^R(\vecr) 
=&\frac{8}{105}\wt{C}_1^2
 \left[
7{\cal P}_0(\mu) \Xi_{\delta\delta,0}^{(0)}(r)
+10{\cal P}_2(\mu) \Xi_{\delta\delta,2}^{(0)}(r)
\right. \nn \\&  \ \ \ \ \ \ \ \ \ \ \ \ \ \ \ \ \ \ \ \ \ \ \left. 
+3{\cal P}_4(\mu) \Xi_{\delta\delta,4}^{(0)}(r)
\right],
\label{eq:ii_la_1d_p} \\
\xi_{-}^R(\vecr) 
=&
\wt{C}_1^2
\left( 1-\mu^2\right)^2  \Xi_{\delta\delta,4}^{(0)}(r)
\nn 
 \\ 
 =&\frac{8}{105}
 \wt{C}_1^2
\left[
7{\cal P}_0(\mu) +10{\cal P}_2(\mu) +3{\cal P}_4(\mu) 
\right]\Xi_{\delta\delta,4}^{(0)}(r).
\label{eq:ii_la_1d_m}
\end{align}
Since the II correlation functions are not affected by RSDs in the LA model, we have $\xi_\pm^S(\vecr) = \xi_\pm^R(\vecr)$. 

Finally, the VI correlation in the LA model in real space is expressed as 
\begin{align}
\xi_{v + }^R(\vecr) 
&=\wt{C}_1
\mu (1-\mu^2)  
\Xi_{\delta\Theta,3}^{(1)}(r). 
\label{eq:vi_la_1d}
\end{align}
The multipole moments of the VI correlation become non-zero only for $\ell=1$ and $\ell=3$ and are given by
\begin{align}
\xi_{v +,1 }^R(r) 
&=-\xi_{v +,3 }^R(r) =
\frac{2}{5}\,\wt{C}_1\,
\Xi_{\delta\Theta,3}^{(1)}(r).
\label{eq:vi_la_1d_l}
\end{align}
Just like the II correlation function, we obtain $\xi_{v+}^S(\vecr)=\xi_{v+}^R(\vecr)$ because RSDs do not affect the VI correlation function in the LA model. 

The linear power spectrum in equation (\ref{eq:xi_formula}) for the LA model is calculated from the \camb code \citep{Lewis:2000}. The LA model breaks down on nonlinear scales and fails to predict the nonlinear smearing of the BAO peak. Thus, as a nonlinear extension of the LA model, we consider to replace the linear power spectrum with its nonlinear counterpart \citep{Bernardeau:2002}. This is called the nonlinear alignment (NLA) model \citep{Bridle:2007} and widely used to model the IA statistics \citep{Blazek:2011,Kirk:2012,Chisari:2013}. As we did in \cite{Okumura:2019}, we use the \regpt code to compute the nonlinear power spectra, $P_{\delta\delta}$ and $P_{\delta\Theta}$, for the NLA model \citep{Taruya:2012,Taruya:2013}. 

\section{$N$-body simulations}\label{sec:simulation}
\subsection{Subhalo catalogs}
As in our earlier studies \citep{Okumura:2017a,Okumura:2018,Okumura:2019}, we use a series of $N$-body simulations of the $\Lambda$CDM cosmology, run as a part of the \darkquest project \citep{Nishimichi:2019}. We employ $n_p=2048^3$ particles of mass $m_p= 8.15875\times 10^{10}\himsun$ in a cubic box of side $L_{\rm box} = 2\higpc$. In total, we have the data set from eight independent realizations and we specifically analyze the snapshots at $z=0.306$.

Subhalos are identified using the  \rockstar algorithm \citep{Behroozi:2013}. The velocity of the subhalo is determined by the average particle velocity within the innermost 10\% of the subhalo radius.  We use the standard definition for the halo mass, $M_h\equiv M_{\rm 200 m}$, defined by a sphere with a radius $R_{\rm 200m}$ within which the enclosed average density is 200 times the mean matter density. We create two subhalo catalogs, one with $M_{h} \geq 10^{13}h^{-1}M_\odot$ and another with $M_{h} \geq 10^{14}h^{-1}M_\odot$. The halo bias is calculated by taking the ratio of the real-space correlation function of the halos to that of the underlying matter measured from our simulations, $b_h^2(r)=\xi_{hh}^R(r)/\xi_{mm}^R(r)$. We find that the linear bias parameters for these catalogs are $b_h=1.65$ and $b_h = 3.11$, respectively.  We assume halos to have triaxial shapes \citep{Jing:2002a} and estimate the orientations of their major axes using the second moments of the distribution of member particles projected onto the celestial plane. 

\begin{figure*}
\includegraphics[width=0.498\textwidth,angle=0,clip]{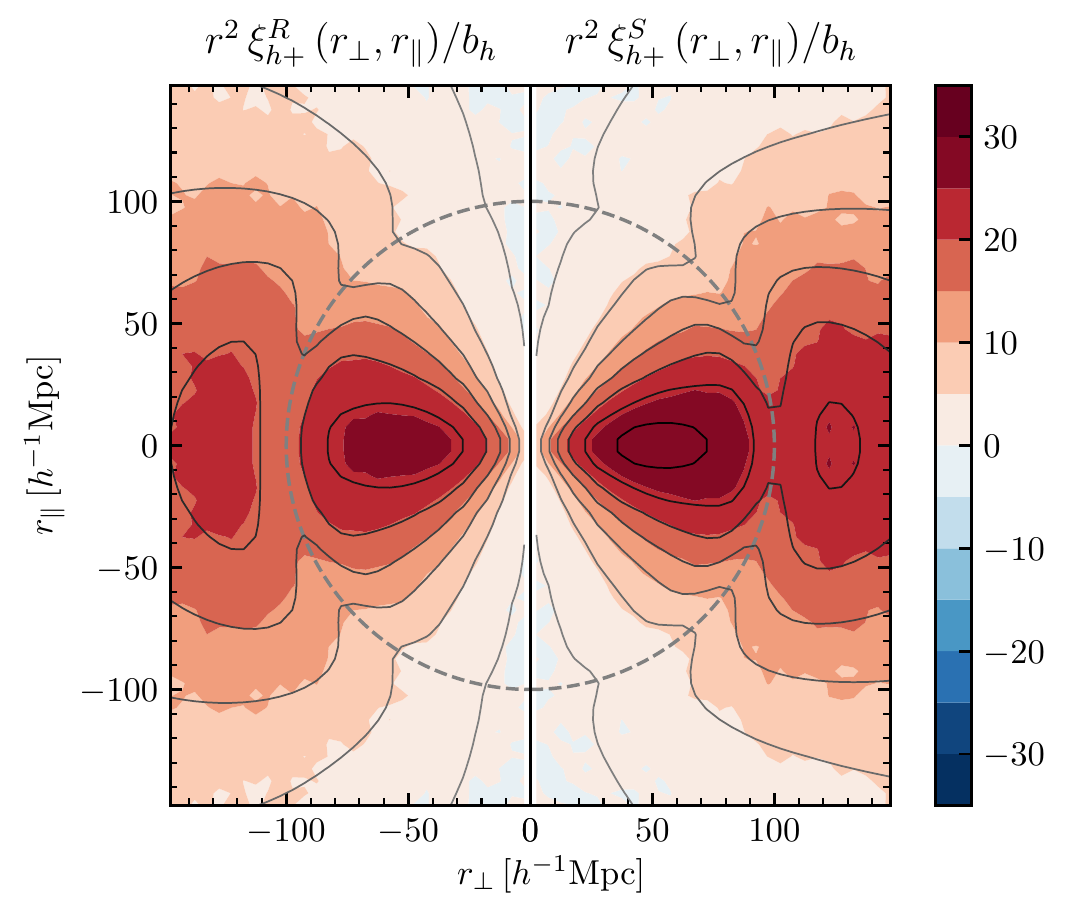}
\includegraphics[width=0.498\textwidth,angle=0,clip]{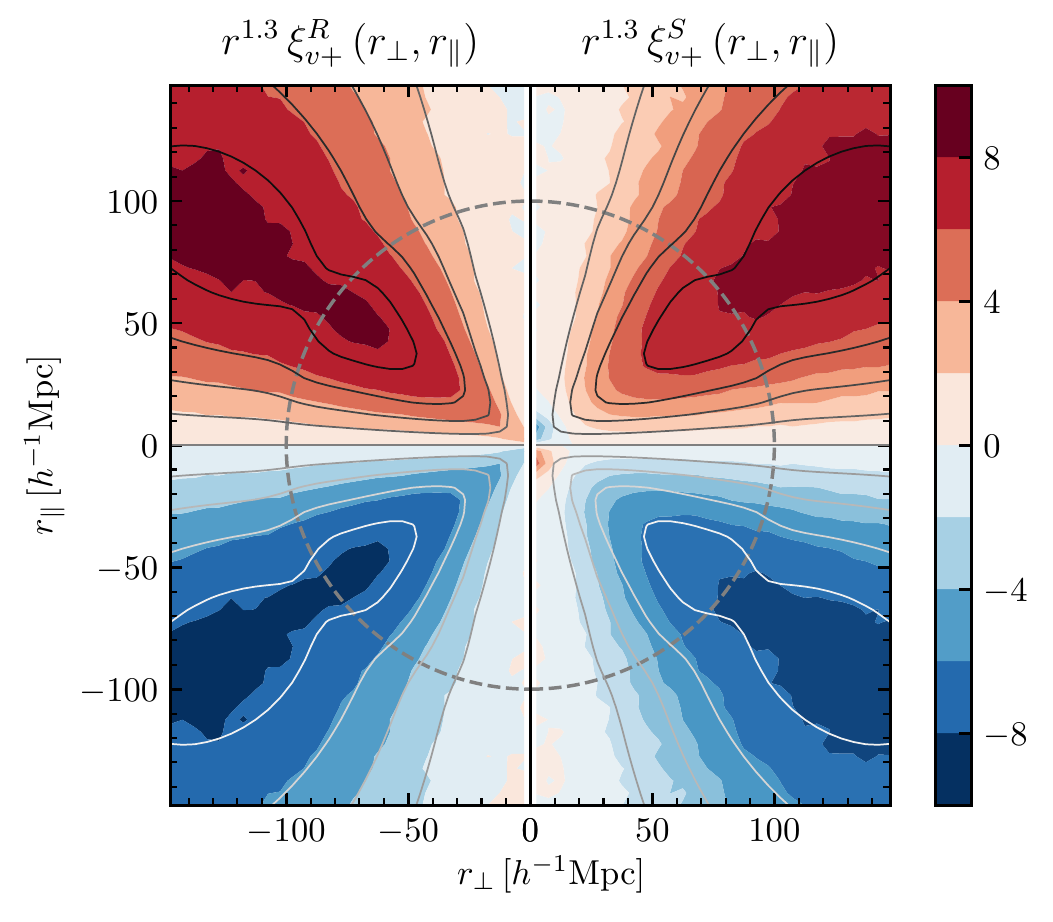}
\includegraphics[width=0.498\textwidth,angle=0,clip]{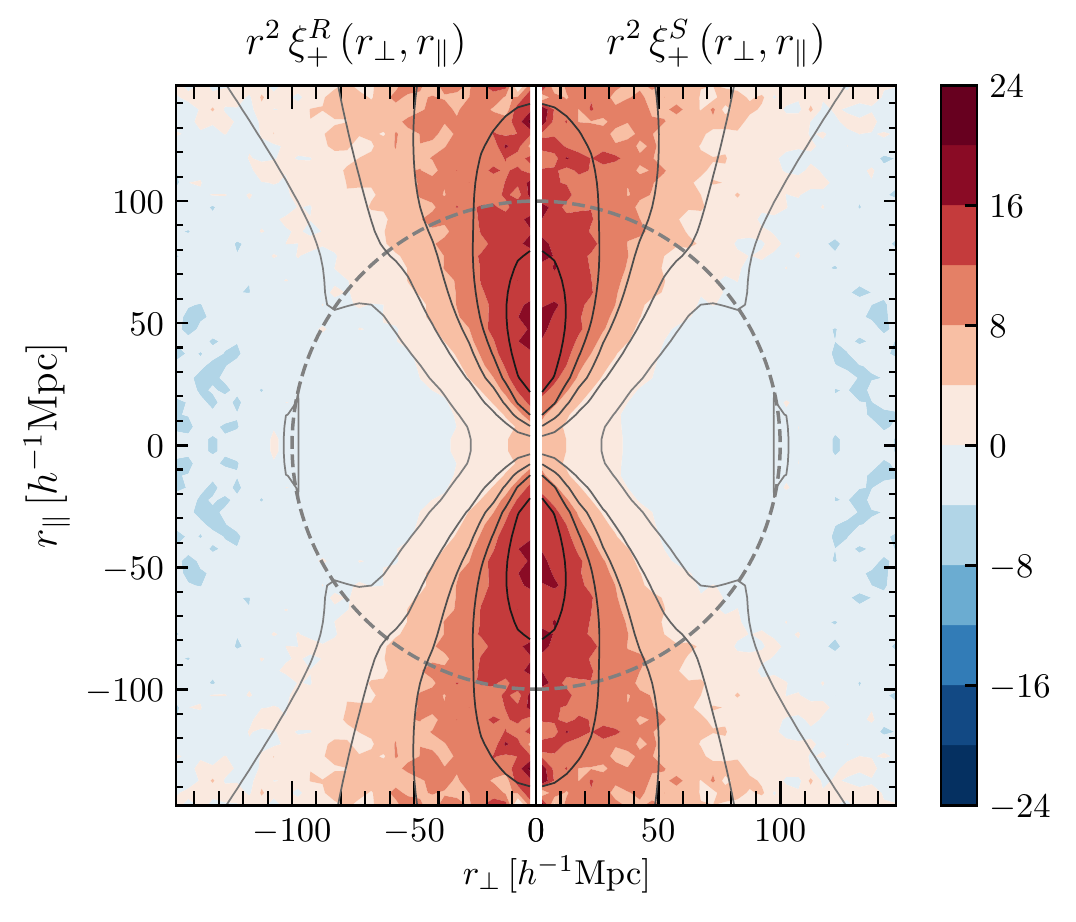}
\includegraphics[width=0.498\textwidth,angle=0,clip]{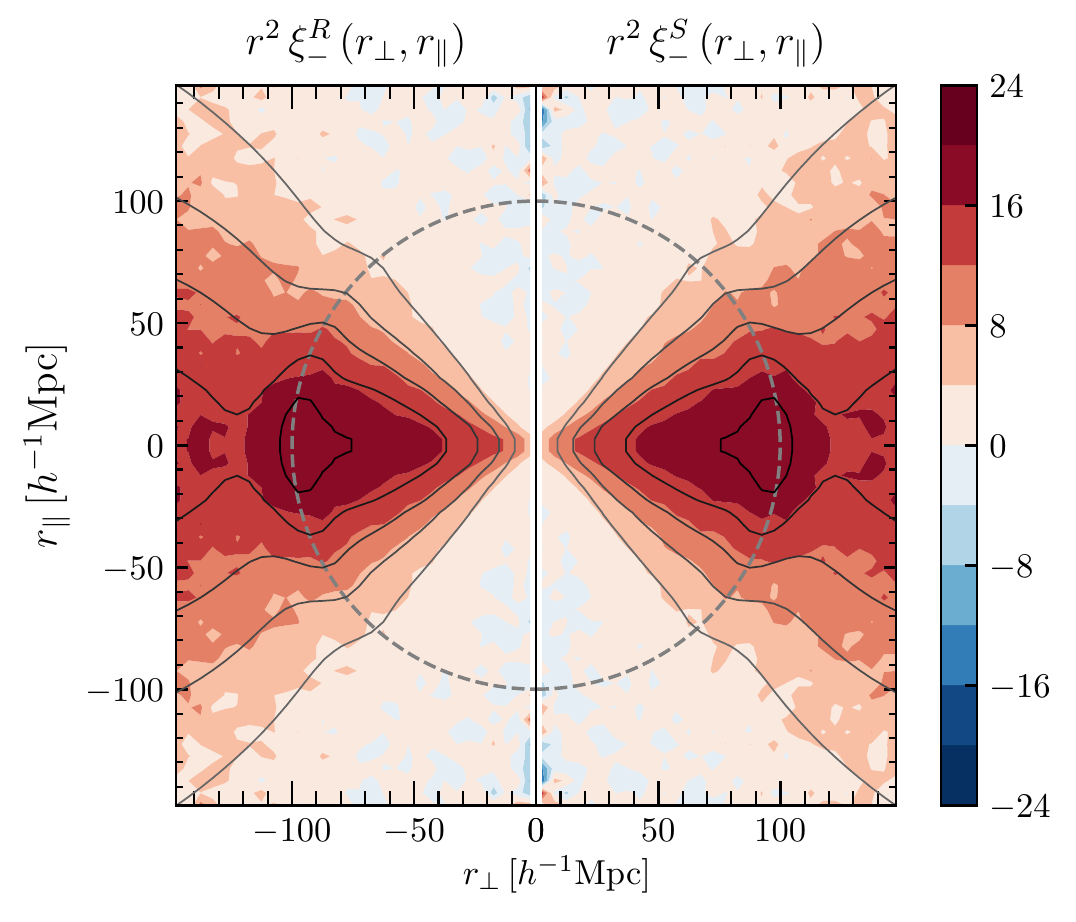}
\caption{
Alignment statistics of subhalos with mass $M_h\geq 10^{13}M_\odot$ as a function of $\vecr=(r_\perp,r_\parallel)$, GI (upper-left), VI (upper-right), and II (lower-left and lower-right) correlations, $\xi_+$ and $\xi_-$, respectively. The left- and right-hand sides of each panel show the statistics in real and redshift space, respectively. In each panel, the color scale shows the measurements from the $N$-body simulations and the gray sold contours show the LA model prediction. The BAO scale, $r\sim 100\himpc$, is denoted by the dashed gray circle. 
}
\label{fig:contour2D_gi_ii_vi_ia_p_rs_2gpc_z014}
\end{figure*}

\begin{figure*}
\includegraphics[width=0.99\textwidth,angle=0,clip]{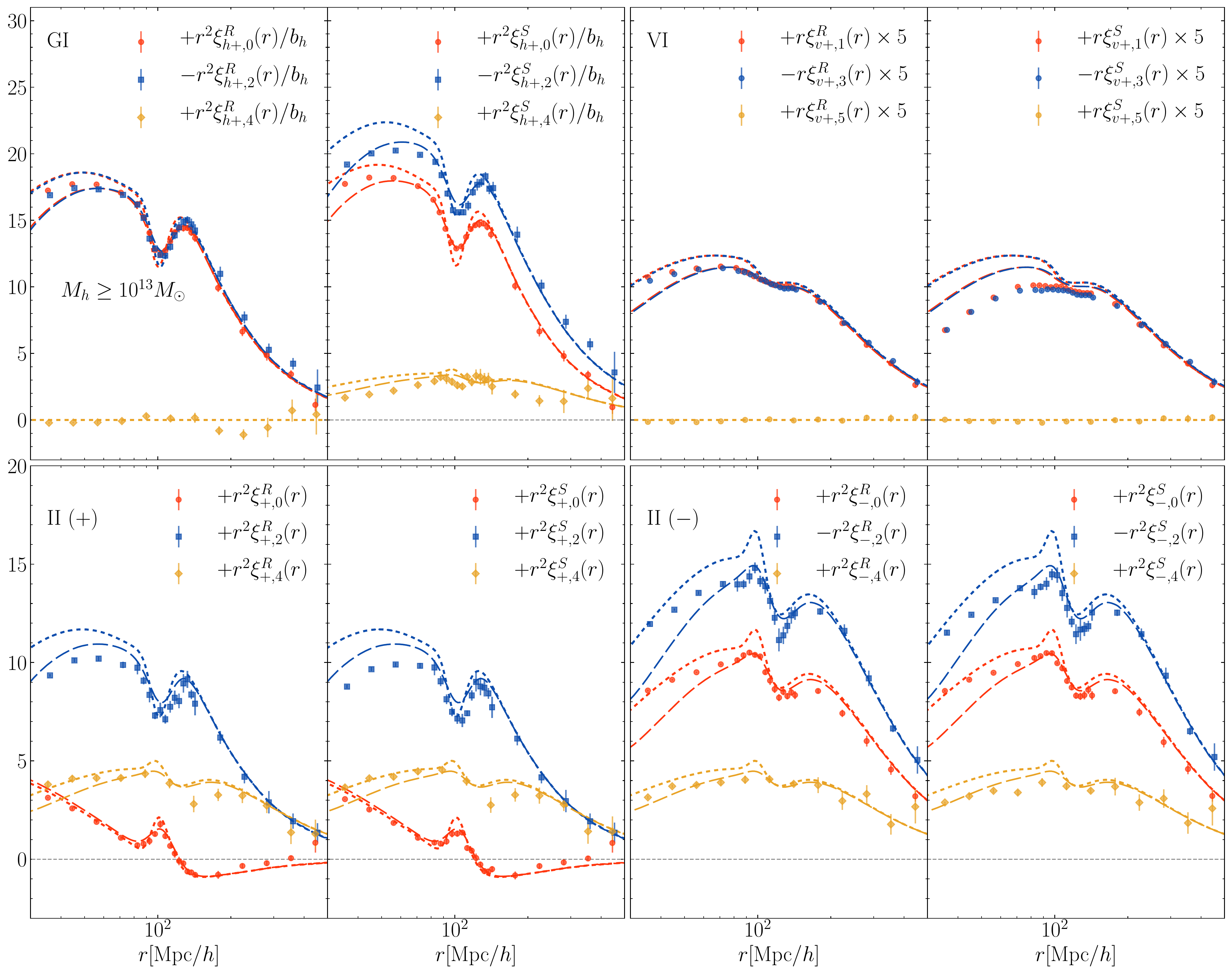}
\caption{Multipole components of alignment statistics of subhalos with mass $M_h\geq 10^{13}M_\odot$, 
$\xi_{h+,\ell}^{(R,S)}$ (upper-left set), $\xi_{v+,\ell}^{(R,S)}$ (upper-right set), $\xi_{+,\ell}^{(R,S)}$ (lower-left set) and $\xi_{-,\ell}^{(R,S)}$ (lower-right set). In each set, the left and right panels show the multipoles in real and redshift space, respectively. While the points show the measurements from $N$-body simulations, the dotted and dashed curves are the corresponding LA and NLA model predictions, respectively. 
}
\label{fig:gi_ii_vi_multi_p_rs_2gpc_z014}
\end{figure*}

\begin{figure}
\includegraphics[width=0.49\textwidth,angle=0,clip]{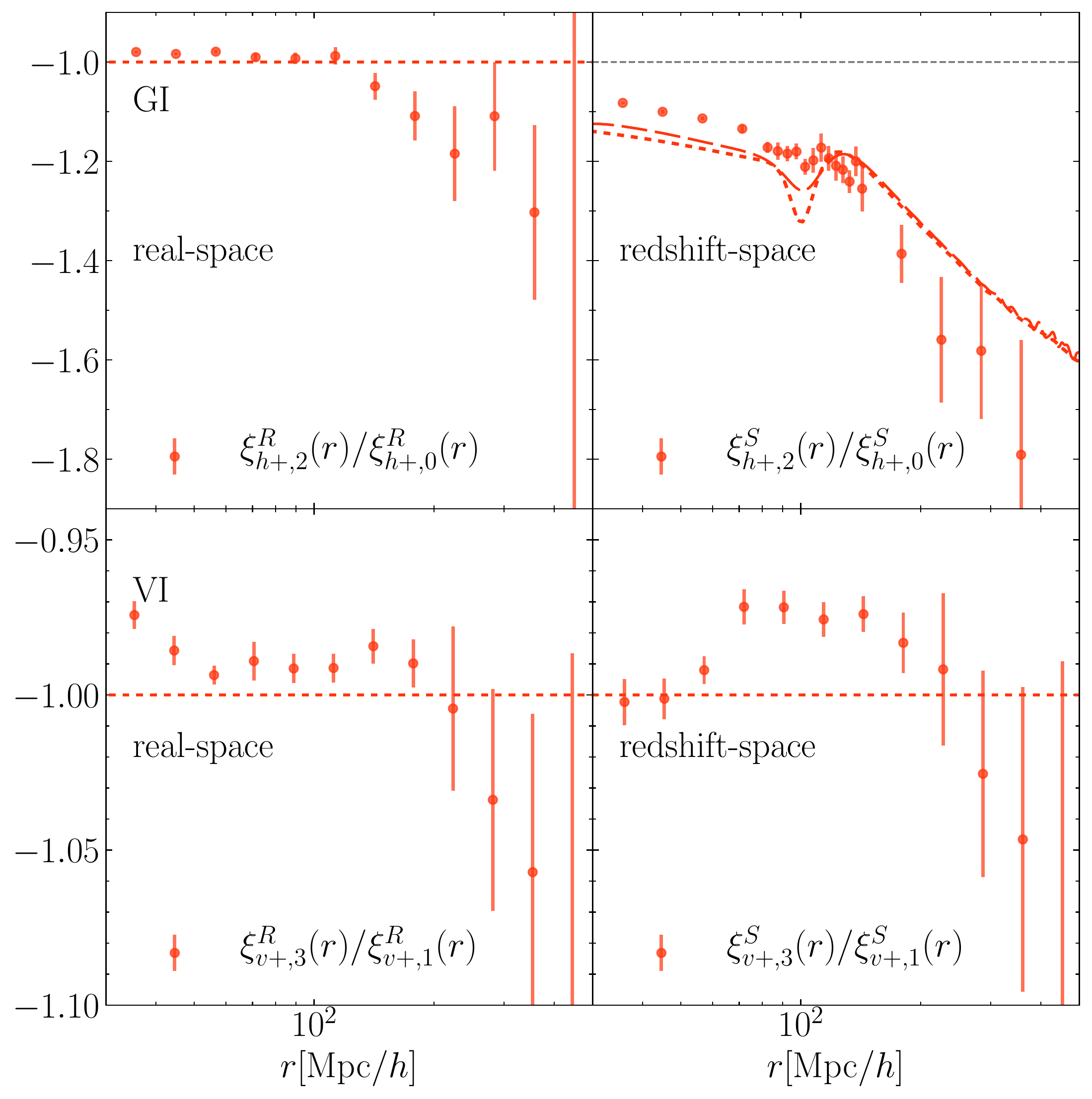}
\caption{Quadrupole-to-monopole ratio of the GI correlation function of suhalos (upper row) and octopole-to-dipole ratio of the VI correlation of subhalos (lower row). The mass range of the subhalo sample is $M_h \geq 10^{13}M_\odot$. In each row, left and right panels show the results in real and redshift space, respectively. The dotted and dashed curves in the upper-right panel are the LA and NLA models, respectively, and in the other panels both the LA and NLA models predict simply $-1$ (red horizontal line). 
}
\label{fig:gi_vi_multi_ratio_p_rs_2gpc_z014}
\end{figure}

\subsection{Estimators}
In this subsection we present estimators to measure from $N$-body simulations the alignment statistics we introduced in Section \ref{sec:definition}. These estimators are the same as those in \cite{Okumura:2019}, but here we extend them to include full angular dependences. We thus measure the statistics in the full 2 dimensional plane, $X(\vecr)=X(r_\perp,r_\parallel)$, as well as their multipole moments $X_\ell(r)$. First, the estimator of the GI correlation function between the density and ellipticity fields is given by
\begin{align}
\xi_{h+}(\vecr) = \frac{1}{RR(\vecr)} \sum_{i,j | \vecr=\vx_j - \vx_i} \gamma_+ (\vx_j), \label{eq:gi_est} 
\end{align}
where $\gamma_+(\vecr)$ is redefined relative to the separation vector $\vecr$ projected on the celestial sphere, and the sum is taken over all pairs with separation $\vecr$. The denominator, $RR(\vecr)$, is the pair count of the random distributions as a function of the vector $\vecr$, which can be analytically and accurately computed because we place the periodic boundary condition on the simulation box. Since we set the number density of the random sample to be equivalent to that of the data so that the pair count does not need to be normalized. The multipole components of the GI function is given by 
\begin{align}
\xi_{h+,\ell}(r) = 
\frac{2\ell+1}{2}\frac{1}{RR(r)}  
\sum_{i,j | r=|\vx_j - \vx_i |} \gamma_+ (\vx_j) {\cal P}_\ell(\mu_{ij}), \label{eq:gi_multi_est} 
\end{align}
where $\mu_{ij}$ is the directional cosine for each pair. Most of the preceding studies multiplied the Legendre polynomial by the angular-binned correlation function, $\xi_{h+}(r,\mu)$. Here we multiply it by each pair before taking the sum in order to obtain more accurate multipoles by avoiding inducing the error due to finite binning. 

The estimator for the VI correlation is similar to that for the GI but given by replacing the density (unity) in equation (\ref{eq:gi_est}) with velocity,
\begin{align}
\xi_{v+}(\vecr) = \frac{1}{RR(\vecr)} \sum_{i,j | \vecr=\vx_j - \vx_i} v_\parallel (\vx_i)\gamma_+ (\vx_j).
\end{align}
Likewise, the estimators for two of the four II correlation functions, $\xi_{(++,\times\times)}$, are given by
\begin{align}
\xi_{++}(\vecr) & = \frac{1}{RR(\vecr)} \sum_{i,j | \vecr=\vx_j - \vx_i} \gamma_+ (\vx_i)\gamma_+ (\vx_j), \\
\xi_{\times\times}(\vecr) & = \frac{1}{RR(\vecr)} \sum_{i,j | \vecr=\vx_j - \vx_i} \gamma_\times (\vx_i)\gamma_\times (\vx_j). 
\end{align}
Then the functions $\xi_\pm$ are obtained through equation (\ref{eq:iipm}). The multipole moments for the II and VI correlation functions are calculated in the same way as that for the GI correlation (equation \ref{eq:gi_multi_est}).

\section{Results}\label{sec:result}

In this section, we make a detailed comparison between the alignment statistics measured from $N$-body simulations and the corresponding model predictions. In Fig.~\ref{fig:contour2D_gi_ii_vi_ia_p_rs_2gpc_z014} we present the GI, II and VI correlation functions for subhalos with $M_h \geq 10^{13}M_\odot$ as a function of $\vecr = (r_\perp,r_\parallel)$ together with the LA model predictions. The left-hand and right-hand sides of each panel shows the real- and redshift-space statistics, respectively. In Fig.~\ref{fig:gi_ii_vi_multi_p_rs_2gpc_z014} we show the multipole moments of these alignment correlation statistics and compared them with the LA/NLA models. In each of four sets, the left and right panels show the real- and redshift-space statistics, respectively. We find $\wt{C}_1/a^2 = 0.94$ for this halo sample with both the LA and NLA models. These are the main result of this paper that we will discuss in more detail in the following subsections. 

\subsection{GI correlation}

The upper-left panel of Fig.~\ref{fig:contour2D_gi_ii_vi_ia_p_rs_2gpc_z014} shows the measurement of the GI correlation function in real and redshift space. Since the GI correlation is an anisotropic function even in real space, it is hard to see the difference between real and redshift space, though the one in redshift space has a larger amplitude. Overall, anisotropic features are well captured by the LA model. The BAO features, which appear as a ring \citep{Matsubara:2004,Okumura:2008,Okumura:2019a}, are also well explained by the model.

The left panel of the upper-left set in Fig.~\ref{fig:gi_ii_vi_multi_p_rs_2gpc_z014} shows the multipole moments of the real-space GI correlation function. The comparison of the monopole measured from the simulations to the LA and NLA models was presented in \cite{Okumura:2019}. Here we extend the comparison to the higher-order multipoles. The tidal alignment models capture the overall shape of the monopole and quadrupole moments of the GI correlation function. A closer look at BAO scales, however, reveals that the LA model starts to deviate from the $N$-body result. Still, the NLA model where the linear power spectra are replaced with the non-linear counterparts gives reasonable agreement even at small scales, $r\sim 50\himpc$. In addition, the smearing of the BAO troughs due to nonlinearities in the monopole and quadrupole moments is accurately predicted by the NLA model. The hexadecapole and higher-order moment measured from $N$-body simulations are consistent with zero, and hence with the LA model. The upper-left panel of Fig.~\ref{fig:gi_vi_multi_ratio_p_rs_2gpc_z014} presents the quadrupole-to-monopole ratio. The $N$-body result becomes $-1$ at all the scales probed in this study, $20\himpc \leq r \leq 500\himpc$. It is predicted by \cite{Okumura:2019a} and our measurement shows that even at the scales where linear perturbation theory breaks down, the LA model still holds. At the large scales, $r\gg 100\himpc$, one can see a small deviation from $-1$ though it is not significant. This deviation is likely to be caused by the finite simulation box. While it is interesting, investigating the effect is beyond the scope of this paper and we leave it for our future work.

The right panel of the upper-left set in Fig.~\ref{fig:gi_ii_vi_multi_p_rs_2gpc_z014} presents the multipoles of the GI correlation function in redshift space, $\xi_{g+,\ell}^S(r)$. The quadrupole in redshift space has a greater amplitude than the monopole. Thus the ratio, $\xi_{g+,2}^S(r) / \xi_{g+,0}^S(r)$, is less than $-1$ due to the effect of RSDs as shown in the upper-right panel of Fig.~\ref{fig:gi_vi_multi_ratio_p_rs_2gpc_z014}. The upper-left panel of Fig.~\ref{fig:gi_ii_vi_multi_p_rs_ratio_2gpc_z014} shows the redshift-to-real-space ratio for the multipoles, $\xi_{g+,\ell}^S(r) / \xi_{g+,\ell}^R(r)$, where $\ell = \{0,2\}$. The enhancement for the GI correlation monopole is only a few percent, as predicted by the red lines \citep{Okumura:2017a}. Rather, the enhancement for the quadrupole moment is more significant, and it is in good agreement with the NLA model at and beyond BAO scales. 

\subsection{II correlation}
\begin{figure}
\includegraphics[width=0.47\textwidth,angle=0,clip]{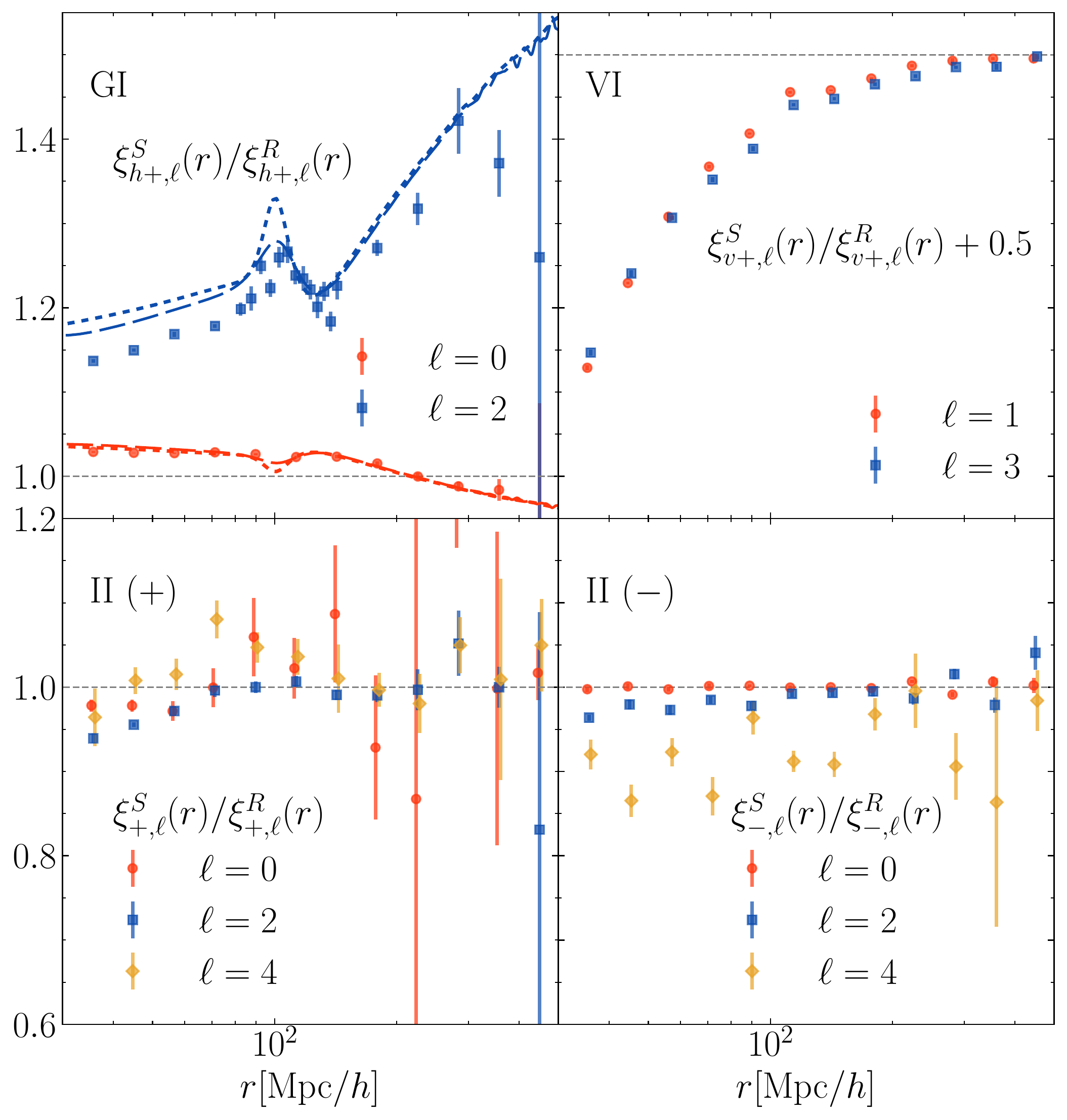}
\caption{
Ratios of IA multipoles in redshift and real space for subhalos with $M\geq 10^{13}M_\odot$. We show those for GI ($\xi_{h+,\ell}^S/\xi_{h+,\ell}^R$) in the upper-left panel, for VI ($\xi_{v+,\ell}^S/\xi_{v+,\ell}^R$) in the upper-right panel, and for II ($\xi_{+,\ell}^S/\xi_{+,\ell}^R$ and $\xi_{-,\ell}^S/\xi_{-,\ell}^R$) in the lower-left and lower-right panels, respectively. The result for the VI correlation function is shifted vertically by $0.5$, so the horizontal dotted line denotes unity.
}
\label{fig:gi_ii_vi_multi_p_rs_ratio_2gpc_z014}
\end{figure}

The lower-left panel of Fig.~\ref{fig:contour2D_gi_ii_vi_ia_p_rs_2gpc_z014} presents the one of the II correlation functions, $\xi_+^{(R,S)}(r_\perp,r_\parallel)$. Overall, we see good agreement between the measurement and the LA model prediction, although it is hard to see the BAO features on this 2-dimensional plane in the $N$-body result. As the LA model predicts, behaviors of the II correlation are not distinguishable in real and redshift space. The lower-left set in Fig.~\ref{fig:gi_ii_vi_multi_p_rs_2gpc_z014} shows the multipole moments of the II correlation functions in real and redshift space, $\xi_{+,\ell}^R(r)$ and $\xi_{+,\ell}^S(r)$, respectively. The BAO signals are clearly detected in the multipoles. The NLA models improve the accuracy for all the non-zero multipole components. Again, there is almost no difference between the II correlations between real and redshift space. It is more clearly demonstrated in the lower-left panel of Fig.~\ref{fig:gi_ii_vi_multi_p_rs_ratio_2gpc_z014}.

In the lower-right panel of Fig.~\ref{fig:contour2D_gi_ii_vi_ia_p_rs_2gpc_z014}, we show another one of the II correlation functions, $\xi_-^{(R,S)}(r_\perp,r_\parallel)$. Just like $\xi_+^{(R,S)}(r_\perp,r_\parallel)$, the $N$-body measurements agree well with the LA model and the measured correlation is almost equivalent between real and redshift space. The multipole moments, $\xi_{-,\ell}^{(R,S)}(r)$, are shown in the lower-right set of Fig.~\ref{fig:gi_ii_vi_multi_p_rs_2gpc_z014}. While both overall shapes of the correlation functions and BAO features are well predicted by the NLA model, the prediction of the LA model systematically deviates from the $N$-body results. The ratio of the correlation function in redshift and real space, $\xi_{-,\ell}^S/\xi_{-,\ell}^R$, is shown in the lower-right panel of Fig.~\ref{fig:gi_ii_vi_multi_p_rs_ratio_2gpc_z014}. Interestingly, while the ratios for the monopole and quadrupole are more or less consistent with unity, that for the hexadecapole deviates from unity by $\sim 10\%$ at all the scales probed. It is partially caused by the nonlinearity of RSDs which cannot be captured by the LA model and beyond the scope of this paper. We will investigate such nonlinearities in future work.

\begin{figure*}
\includegraphics[width=0.99\textwidth,angle=0,clip]{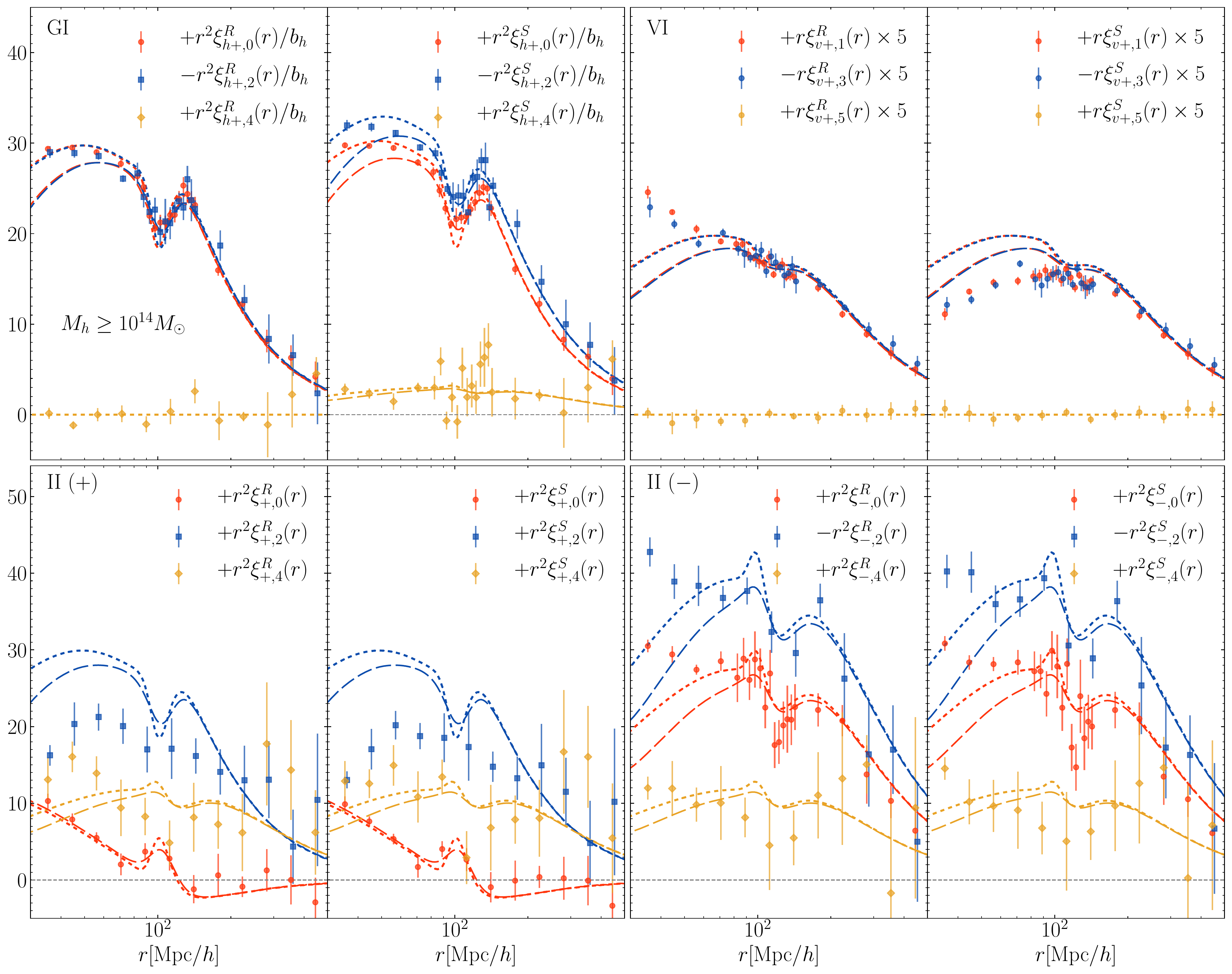}
\caption{Same as Fig. \ref{fig:gi_ii_vi_multi_p_rs_2gpc_z014}, but for 
the alignment statistics with more massive halos, $M\geq 10^{14}M_\odot$. 
}
\label{fig:gi_ii_vi_multi_c_rs_2gpc_z014}
\end{figure*}

\subsection{VI correlation}
The VI correlation function of subhalos is shown as a function of $\vecr=(r_\perp,r_\parallel)$ in the upper-right panel of Fig.~\ref{fig:contour2D_gi_ii_vi_ia_p_rs_2gpc_z014}. Again, the difference between the measurements in real and redshift space is small. However the agreement with the LA model gets worse in redshift space than in real space, as expected. Since the VI correlation function depends on odd powers of $\mu$, the sign of the function flips for $r_\parallel >0$ and $r_\parallel <0$. Moreover, due to the nonlinear RSD called the finger-of-god effect, the sign of the VI correlation is further changed at $r<10\himpc$ \citep[see, e.g.,][]{Okumura:2014}. 

The multipoles of the VI correlation function in real space are shown in the left-hand side of the upper-right set of Fig.~\ref{fig:gi_ii_vi_multi_p_rs_2gpc_z014}. The real-space VI dipole has been already presented in \cite{Okumura:2019}. The octopole measured from the simulations shows a behavior very similar to the dipole. The octopole-to-dipole ratio of the VI correlation in real space is shown in the lower-left panel of Fig.~\ref{fig:gi_vi_multi_ratio_p_rs_2gpc_z014}. Although the measured VI multipoles start to deviate from the NLA model at $r \sim 60\himpc$, the octopole-to-dipole ratio is consistent with the prediction of the tidal alignment model, $-1$, within 1\% to slightly smaller scales.

The multipoles of the VI correlation function in redshift space are significantly suppressed, even at BAO scales, as shown in the right-hand side of the upper-right set of Fig.~\ref{fig:gi_ii_vi_multi_p_rs_2gpc_z014}. The BAO features detected in real space are smeared out in redshift space. Still, the octopole-to-dipole ratio is consistent with $-1$, as predicted by the LA/NLA models (the lower-left panel of Fig.~\ref{fig:gi_vi_multi_ratio_p_rs_2gpc_z014}), though the accuracy gets slightly worse, to $\sim 3\%$. The ratios of the VI correlation multipoles in redshift and real space are shown in the upper-right panel of Fig.~\ref{fig:gi_ii_vi_multi_p_rs_ratio_2gpc_z014}. We clearly see the suppression of the redshift-space correlation at small scales. The suppression of the VI correlation is due to the nonlinear RSDs, and it reaches $\sim 40\%$ at $r=30\himpc$. It qualitatively makes sense because the density-weighted velocities are known to be significantly affected by the finger-of-god (FoG) effect \citep{Okumura:2014}. However, the FoG effect appeared at scales much larger than we expected. We will investigate it using nonlinear perturbation theory in future work. 

\begin{figure}
\includegraphics[width=0.47\textwidth,angle=0,clip]{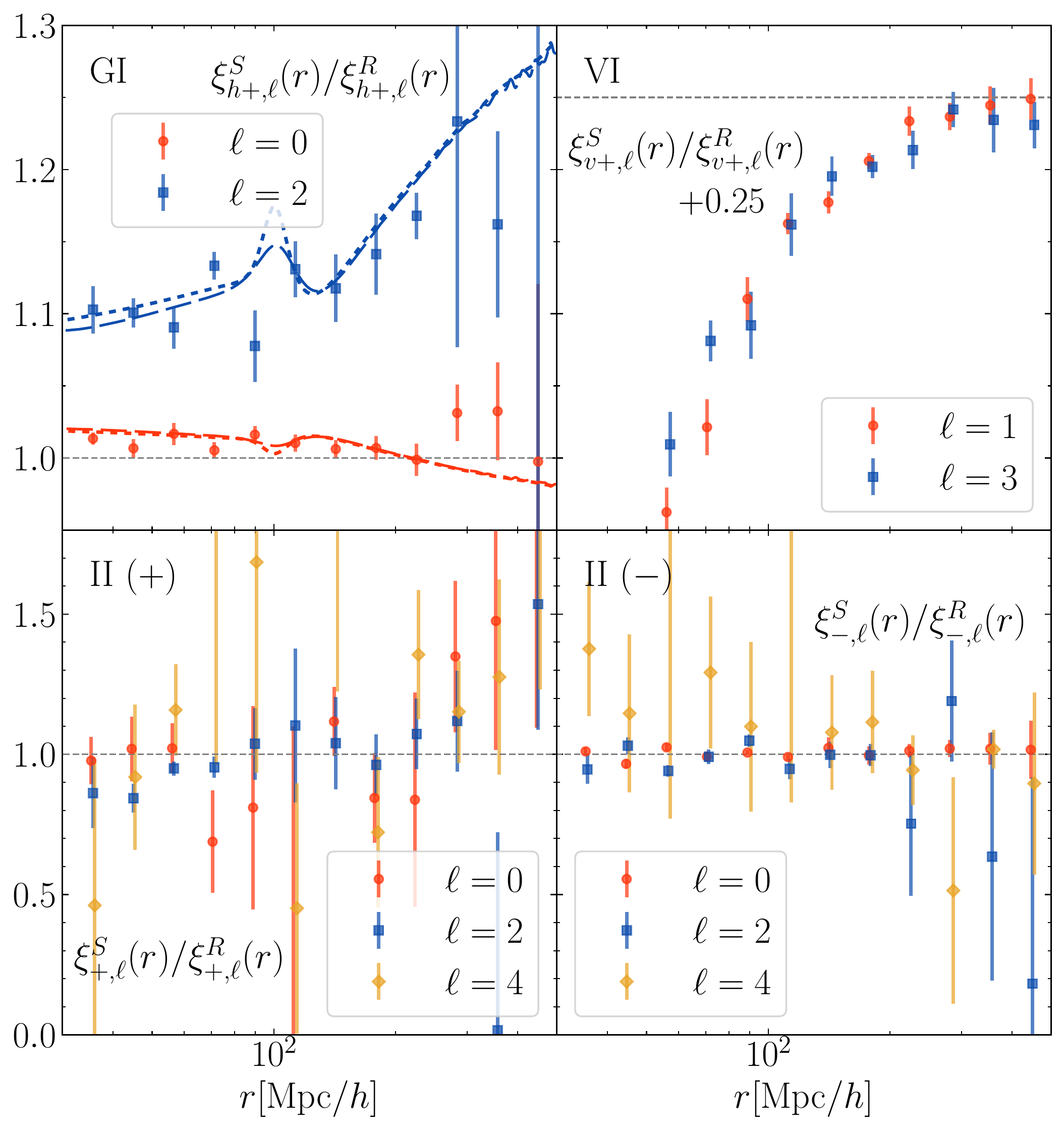}
\caption{
Same as Fig. \ref{fig:gi_ii_vi_multi_c_rs_ratio_2gpc_z014}, but for more massive halos with $M\geq 10^{14}M_\odot$.
}
\label{fig:gi_ii_vi_multi_c_rs_ratio_2gpc_z014}
\end{figure}

\subsection{Halo mass dependence of IA}

So far, we have analyzed only one subhalo sample, with mass $M_h \geq 10^{13}M_\odot$. It is, however, well known that the amplitude of IAs strongly depends on the halo mass \citep{Jing:2002} \citep[see also][for recent studies]{Xia:2017,Piras:2018}. Thus, we analyze a more massive subhalo catalog, with mass $M_h \geq 10^{14}M_\odot$, and repeat the above analysis. Since except for the amplitude, behaviors of the alignment statistics are more or less the same as the results presented so far, we will show only the results of the multipoles correlation functions, not those of the 2-dimensional correlation functions. 

First, we find that the parameter of the IA amplitude is $\wt{C}_1/a^2 = 1.50$. Compared to the subhalo sample with $M_h \geq 10^{13}M_\odot$, the amplitude is increased by a factor of $1.6$ and it is less significant than the bias (a factor of $1.9$ enhancement). Fig.~\ref{fig:gi_ii_vi_multi_c_rs_2gpc_z014} shows the multipole correlation functions in real and redshift space. The monopole of the GI correlation function in real space has been presented in \cite{Okumura:2019} for this massive halo sample. The nonlinearities of the GI correlation are slightly more significant than those in the less massive halos in both real and redshift space. As already seen in \cite{Okumura:2019}, the BAO features in the GI function are more significant than the NLA model prediction because in peak theory the BAO features are amplified for higher peaks \citep{Desjacques:2008}. The upper-left panel of Fig.~\ref{fig:gi_ii_vi_multi_c_rs_ratio_2gpc_z014} shows the ratios of the GI multipole correlation functions in redshift and real space, $\xi_{h+,\ell}^S/\xi_{h+,\ell}^R$. The enhancement of the GI correlation functions due to RSDs is suppressed for massive halos by the RSD parameter, $f/b_h$. 

Next, one of the two II correlations, $\xi_{+,\ell}^{(R,S)}$, for halos with mass $M_h\geq 10^{14}M_\odot$ is shown in the lower-left set of Fig.~\ref{fig:gi_ii_vi_multi_c_rs_2gpc_z014}. Although the amplitude becomes larger due to the factor of $\wt{C}_1^2$, the measurement itself becomes much noisier. We thus do not focus on the BAO features although they are detected by all the multipole components of $\xi_{+,\ell}^{(R,S)}$. The ratios of the II correlations in redshift and real space are shown in the lower-left panel of Fig.~\ref{fig:gi_ii_vi_multi_c_rs_ratio_2gpc_z014}. Again, while the measurements are noisy, they are consistent with unity. Another of the two II correlations, $\xi_{-,\ell}^{(R,S)}$, is shown in the lower-right set of Fig.~\ref{fig:gi_ii_vi_multi_c_rs_2gpc_z014}. Due to the fact that the ellipticity is the density-weighted field, the measurements significantly deviate from the LA and NLA models compared to the case of lower-mass halos. However, the ratios of the correlation functions in redshift and real space are consistent with unity even on small scales, in agreement with the LA/NLA models. 

Finally, the multipoles of the VI correlation function, $\xi_{v+,\ell}^{(R,S)}$ are presented in the upper-right set of Fig.~\ref{fig:gi_ii_vi_multi_c_rs_2gpc_z014}. Since just like the II correlation functions, both the measured ellipticity and velocity fields are density-weighted, the VI correlation function is significantly affected by the nonlinearities. On the other hand, the VI correlation function is severely suppressed in redshift space. As shown in the upper-right panel of Fig.~\ref{fig:gi_ii_vi_multi_c_rs_ratio_2gpc_z014}, the ratios of the dipole and octopole in redshift space to those in real space are more or less equivalent, and the behaviors are also the same as those for less massive halos. Although the nonlinear behaviors of the II and VI correlation functions are interesting, understanding these effects are beyond the scope of this paper and we will investigate them in future work. 

\section{Conclusions}\label{sec:conclusion}
The intrinsic alignment (IA) of galaxy images can be utilized as a powerful cosmological probe. However, whether or not the IA statistics can be useful for this purpose entirely depends on the accuracy of the theoretical modeling of the relevant statistics. In this paper, we test analytical predictions of the tidal alignment model for IA statistics derived in \cite{Okumura:2019a}, using a large set of cosmological $N$-body simulations. We measured various alignment statistics, the GI, II and VI correlation functions, in real and redshift space with the full 2-dimensional plane (Fig.~\ref{fig:contour2D_gi_ii_vi_ia_p_rs_2gpc_z014}) as well as with the multipole expansions (Fig.~\ref{fig:gi_ii_vi_multi_p_rs_2gpc_z014}). We find that both anisotropies of BAOs and RSDs are accurately predicted by the linear alignment (LA) model. This demonstrates that the IA encoded in the large-scale structure of the universe can be used as geometric and dynamical probes to constrain cosmological parameters.

In a companion paper, \cite{Taruya:2019}, we considered the observations of the BOSS and DESI galaxy surveys and indeed showed that combining IA statistics to the conventional galaxy clustering statistics allows us to significantly tighten the cosmological parameter constraints. IAs of galaxies in the BOSS surveys have already been detected \citep{Li:2013,Singh:2015,van-Uitert:2017}. However, all these studies focused on the angular-averaged components of the GI and II correlation functions. We will present the cosmological analysis of IA statistics in redshift space using the higher-order multipoles in future work. 

In order to extract more information from the IA statistics, one needs to develop a model at quasi non-linear scales. In this paper we presented the non-linear alignment model, where the linear power spectra in the LA model are replaced by the nonlinear power spectra. However, this model is simply an approximation and does not provide a consistent nonlinear description. Recently there have been attempts to model IAs of galaxies based on nonlinear perturbation theory beyond the LA model \citep[see e.g.,][]{Blazek:2019,Vlah:2019}. The detailed comparison of such nonlinear models of IA statistics to the $N$-body simulation results will be performed in future work. 

\section*{Acknowledgments}
We thank Aniket Agrawal and Yin Li for useful discussion. T.~O. acknowledges support from the Ministry of Science and Technology of Taiwan under Grants No. MOST 106-2119-M-001-031-MY3 and the Career Development Award, Academia Sinina (AS-CDA-108-M02) for the period of 2019 to 2023. A.~T. was supported in part by MEXT/JSPS KAKENHI Grants No. JP15H05889 and No. JP16H03977. T.~N. was supported by Japan Science and Technology Agency CREST JPMHCR1414, and by JSPS KAKENHI Grant Number JP17K14273 and JP19H00677. This research was supported by the World Premier International Research Center Initiative (WPI), MEXT, Japan. Numerical computations were carried out on Cray XC30 and XC50 at the Center for Computational Astrophysics, National Astronomical Observatory of Japan.

\bibliographystyle{mnras} 

\label{lastpage}
\end{document}